\documentclass[]{aiaa-tc}

 \title{ Prototype VOEvent Network Systems
         based on VTP and XMPP for the SVOM Chinese Science Center}

 \author{
  Mo Zhang%
    \thanks{Engineer, Research Division of Space Science, NAOC, 20A Datun Road, Chaoyang District, Beijing}
  \ Maohai Huang%
  \thanks{Professor, Research Division of Space Science, NAOC, 20A Datun Road, Chaoyang District, Beijing}
     \ and Chao Wu%
   \thanks{Associate Professor, Research Division of Space Science, NAOC, 20A Datun Road, Chaoyang District, Beijing}\\
  {\normalsize\itshape
   Key Laboratory of Space Astronomy and Technology, National Astronomical Observatories,
   }\\
    {\normalsize\itshape
    Chinese Academy of Sciences, Beijing, 100012, China}
 }

 \AIAApapernumber{YEAR-NUMBER}
 \AIAAconference{Conference Name, Date, and Location}
 \AIAAcopyright{\AIAAcopyrightD{YEAR}}

\begin{document}
\bibliographystyle{plain}
\maketitle

\begin{abstract}
We present the current progress of design and build of two prototype VOEvent network systems for the SVOM Chinese Science Center. One is based on VTP which is compatible with the global VOEvent network, the other is based on XMPP which enables cross-platform messaging and information sharing among human users. We also present a demonstration of VOEvent controlled follow-up observation, including triggering, observational data transferring, as well as other procedures.
\end{abstract}

\section{Background}
 \label{s:intro}
Time domain astronomy is devoted to the detection and study of rare time-varying phenomena in the universe, including catastrophic outbreak (such as supernovae and gamma-ray bursts (GRBs)), intense energy release phenomena (such as Novas and variability of AGNs), etc. During a typical observation for such phenomena, when a transient is detected, due to limited observation time and wavelength regimes of a individual telescope, an transient alert is raised and disseminated to other telescopes through a dedicated event network to do fast and comprehensive follow-up observations. For many types of transient events, astronomical knowledge is gained through such follow-up observations~\cite{2006ASPC..351..637W}.


In order to make it possible for automatically parsing transient events, the IVOA (International Virtual Observatory Alliance) established a VOEvent (Virtual Observatory Event) specification which defines the format of various information types of observed transient events, such as location, time, etc. Information about each event is encapsulated in a structured XML (Extensible Markup Language) file and disseminated through a ground-based VOEvent network. Complying with this VOEvent format standard, several event forwarding and archiving data centers such as GCN (Gamma-ray Coordinates Network)\footnote{http://gcn.gsfc.nasa.gov/} and 4PiSky\footnote{http://4pisky.org/}, have been established. These servers are connected, forming a timely and reliable global VOEvent network with other follow-up telescopes, and are serving for astronomical space missions that are in operation (such as SWIFT, FERMI, etc.~\cite{2012IAUS..285...41G}).

The SVOM (Space-based multi-band astronomical Variable Object Monitor) is a Chinese-French space mission dedicated to the detection, localization, and study of GRBs and other high-energy transient phenomena\cite{2011CRPhy..12..298P}. SVOM is planned to be launched in 2021, with a life of 3--5 years. According to data flow design of SVOM mission, the data from onboard instrument are first sent to the FSC (French Science Center) and then forwarded to the CSC (Chinese Science Center). Meanwhile, FSC and CSC are responsible for forwarding the VOEvents to their GFTs (Ground Follow-up Telescopes) respectively. Therefore, it is essential to build a well designed VOEvent network in the development of the SVOM ground segment.



%

As far as SVOM is concerned, besides building a network based on VTP\cite{protocol} (VOEvent Transport Protocol), we can build a network based on XMPP (Extensible Messaging and Presence Protocol) as an alternative. Comprising with VTP, XMPP is broadly used in different areas, well specified in a bunch of RFC (references)\footnote{http://xmpp.org/rfcs/}, XEP (extensions)\footnote{http://xmpp.org/extensions/} as well as with a lot of tools. Since the decision has not been made yet, we planned to build and test a prototype VOEvent network system based on each protocol, so as to experiment different technologies, and to accumulate experiences for future development.




In chapter~\ref{s:design}, we introduce the design of the two prototypes. In chapter~\ref{s:case} we present a demonstration to illustrate the design details of the prototype based on XMPP. At last, in chapter~\ref{s:plan} we give the future plan of the SVOM VOEvent network system.


\section{Prototype Design}
 \label{s:design}

\subsection{Prototype network system based on VTP}


VTP is a protocol for VOEvent transportation which has been in use by nodes in the VOEvent distribution network such as GCN and 4PiSky for several years. Our prototype VOEvent network is based on Comet\cite{2014A&C.....7...12S}, an open source software implementation of VTP. Comet provides a mechanism for fast and reliable VOEvent distribution. It implements all functionalities for every roles in a VOEvent network, such as the publisher, subscriber or filter, making the network easy to build and maintain. We can use Comet to subscribe to VOEvent streams, publish our own events to the community, or select only events that we are interested in.

 \begin{figure}[h!]
 \centering
 \includegraphics[width=0.8\textwidth]{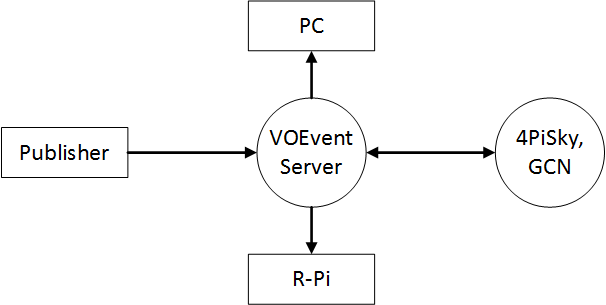}
 \caption{Prototype VOEvent network topology based on VTP}
 \label{f:vtp}
\end{figure}

The topology of the prototype is shown in Figure~\ref{f:vtp}. It includes a main broker which is responsible for handling event transactions not only with SVOM CSC users but also with external brokers. In addition to setting up a subscriber client on a PC, we also configured a client running on a Raspberry Pi, which is a low cost, credit-card sized computer based on ARM.  This prototype network has been in operation for several months, continuous receiving and storing events from the GCN. We also contacted 4PiSky to add the IP address of our broker to its whitelist and we have successfully sent some test VOEvents to it. We hope to make our broker become a part of the global VOEvent network, so that SVOM can benefit from other space astronomical missions, and can also disseminate VOEvents produced by itself as well as its GFTs to the global VOEvent community.

It is worth noticing that the network system is very scalable and flexible due to the tree topology and the software consistence. Besides, Raspberry Pi is very suitable to serve as brokers or filters in the network to select events of different science cases. This two features enable us to easily expand the prototype to a more powerful and structured VOEvent network.


\subsection{Prototype network design based on XMPP}

Compared with VTP, there are various alternatives cross-platform clients based on XMPP, even on mobile devices. Besides, the payload on a XMPP network is not limited to VOEvents, but also can be in other formats like text messages and pictures. Given these advantages, we carefully design the network structure to satisfy various use cases of different roles.

\begin{figure}[htb]
  \centering
 \includegraphics[width=0.8\textwidth]{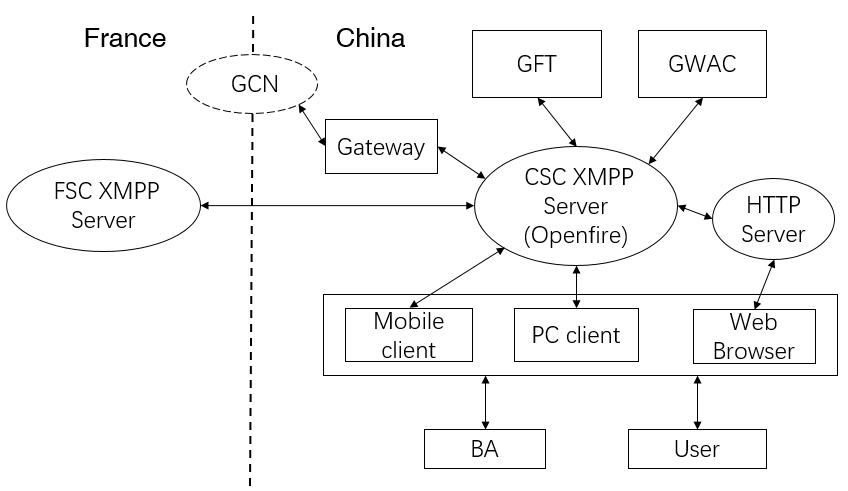}
 \caption{VOEvent network topology for SVOM CSC}
 \label{f:cfxmpp}
\end{figure}

We use Openfire on the server side of the prototype network, which is an instant messaging and group chat server based on XMPP. Openfire is designed to be a centralized system that serves many clients. As shown in Figure~\ref{f:cfxmpp}, it not only forward VOEvents to telescopes such as GFT and GWAC\footnote{Ground Wide Angle Cameras, a payload of SVOM.},  but also handles interactions between a variety of cross-platform clients, as well as web browsers through an HTTP server. BAs (burst advocates) and users can chat in natural language and sharing information between each other. BA also has the privilege to send VOEvents to telescopes.

 \begin{figure}[htb]
 \centering
 \includegraphics[width=0.8\textwidth]{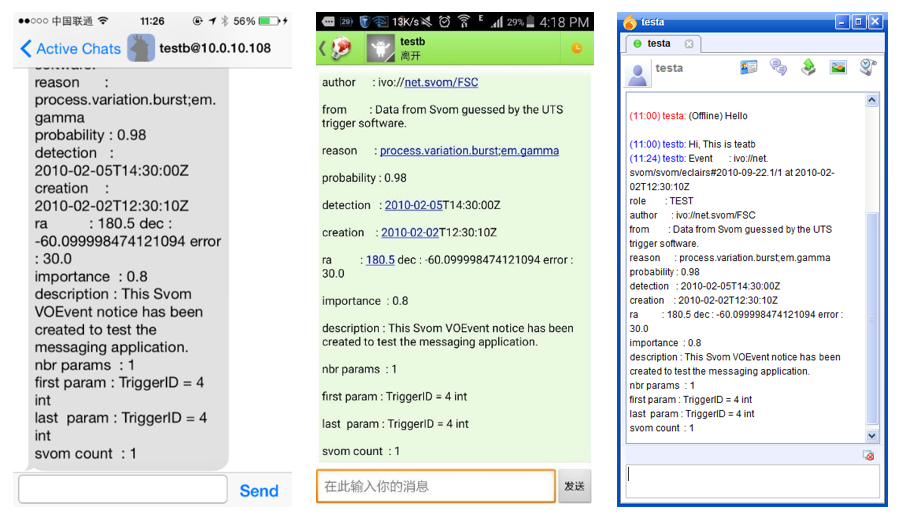}
 \caption{We have tested sending VOEvents to three different clients based on XMPP, including an iOS client  ��Monal�� (left), an Android client ��Xabber�� (middle) and a PC client ��Spark�� (right).}
 \label{f:mobiles}
\end{figure}

Involving various of cross-platform clients is a highlight of the prototype system, which greatly expands the usage scenarios for different user roles. We have tested sending some parsed alert messages to three of them, as shown in Figure~\ref{f:mobiles}.

For SVOM, based on the data flow we have mentioned in Chapter~\ref{s:intro}, the network is considered to include two main servers, one locates in FSC and the other locates in CSC. Since the communication interface between the two XMPP servers has not been well discussed, we only show the network topology design for the Chinese side. When the inter-server connection between CSC and FSC is established, this design concept can be easily transplanted to the French side.


\section{A demonstration of Follow-up Observation Workflow based on XMPP}
 \label{s:case}
In the demonstration, VOEvent serves as containers that carries four types of information: trigger, observational results, data upload status from GFT to CSC, and receipt. We configured a VOEvent module on each side. Besides, there are two more modules in the system. One is called CSC simulator which handles triggering and data archiving; the other is called GFT simulator which handles a batch of simulation observation processes.


\begin{figure}[htb]
\centering
\includegraphics[width=1.0\textwidth]{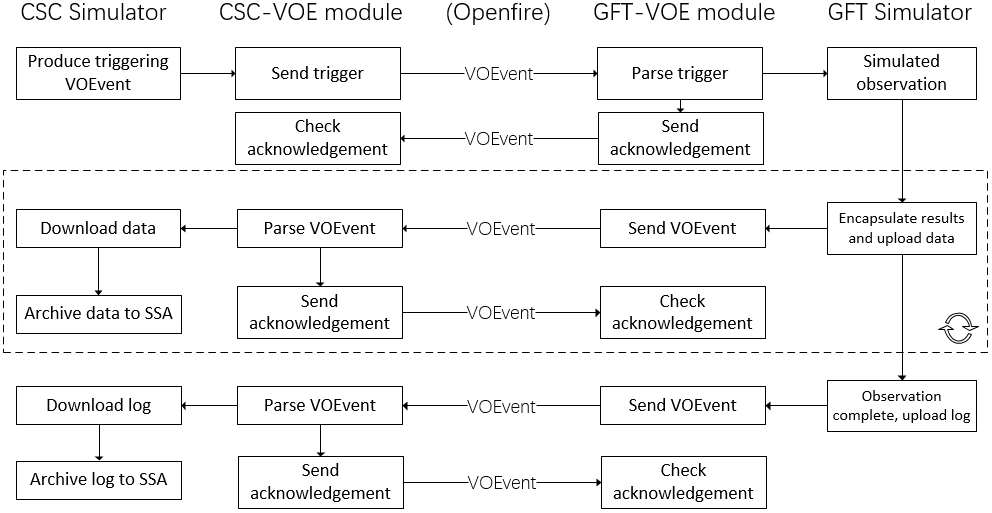}
\caption{VOEvent workflow between SVOM CSC and GFT}
\label{f:task}
\end{figure}

The details of the process is as follows: first, a trigger event is produced and sent from CSC to GFT, then GFT parse out the trigger information and followed by a process of follow-up observations. A follow-up observation may contain several tasks that scheduled at different times. As the observation continues, some important results may be produced and updated, these results such as SED (spectral energy distribution) and light curve, will be formatted and encapsulated into a VOEvent packet and then sent to CSC. When an observation task completed, GFT uploads all observational data to the FTP server at CSC. In addition, it sends a VOEvent right before and after the transmission separately to notify CSC the data upload status, so that CSC can download from the FTP server and archive data into SSA (SVOM science archive) as soon as possible. After all observation tasks are accomplished, GFT uploads the processing logs and notify CSC. Please note that each sending of a VOEvent packet needs a receipt to ensure it is successfully delivered, see Figure~\ref{f:task}.

 \begin{figure}[htb]
 \centering
 \includegraphics[width=0.8\textwidth]{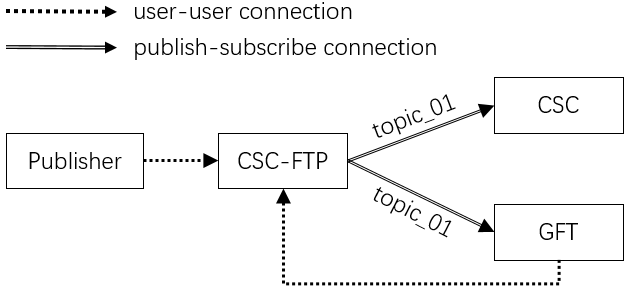}
 \caption{ Inter-module VOEvent connection types (all connections are via Openfire)}
 \label{f:linktype}
\end{figure}

To achieve the automatic process, we deploy VOEvent modules at CSC, CSC-FTP and GFT separately. On each module, we install a JAVA software developed by the our development team to handle VOEvent packets based on XMPP. We configure these modules to register on the Openfire server as different users and to adopt different actions when receiving different types of packet. The inter-module connection types are shown in Figure~\ref{f:linktype}. Connections for triggering and for events from GFT use the user-user connection, while CSC and GFT subscribe the same topic that created by the CSC-FTP. Please note that GFT cannot talk to CSC directly, the message is relayed by CSC-FTP. This design model is consistence with the data flow. By far this system works well but could be improved along with the iterative development.



%

\section{Future Plan}
 \label{s:plan}
We will make improvement for the design of the prototypes and to evolve it to a complete VOEvent system before the launch of SVOM. Currently we plan to deploy the XMPP prototype on one of the SVOM GFTs in China, so that an automated follow-up observation can be conducted. This process starts from triggering, followed by GFT making observations and data reductions on-site, until CSC finally archiving all data and associated VOEvents to the database. Besides, as shown in Fig~\ref{f:cfxmpp}, there are some interfaces that still need to be implemented and tested, including the inter-server VOEvent communication between CSC and FSC, web-based and Email-based VOEvent publishing. We are also have a plan to develop an interface software between VTP and XMPP to take advantages of both protocols, in order to make the network system not only compatible with the global VOEvent network, but also support cross-platform information sharing. Introducing XMPP protocol is a major innovation for the SVOM VOEvent network. We expect this innovation be more widely applied by other space missions in the fields of time domain astronomy.

\section*{Acknowledgments}
This work was supported by National Natural Science Foundation of China (No.11603049) and a strategic pilot science and technology programme for space science, Chinese Academy of Sciences (Grant Number: XDA040803).

\bibliography{bibtex_database}


\end{document}